\documentclass[letterpaper, 10 pt, conference]{ieeeconf}  % Comment this line out if you need a4paper
\usepackage{graphicx}
\usepackage{epstopdf}
\usepackage{color}
\usepackage{tabularx}
\usepackage{caption}
\usepackage{multirow}
\usepackage{booktabs} 
\usepackage{subcaption}
\usepackage{float}
\usepackage[table]{xcolor}
 \usepackage{url}

\UseRawInputEncoding
\IEEEoverridecommandlockouts                              % This command is only needed if 
\title{\LARGE \bf {Privacy Perceptions in Robot-Assisted Well-Being Coaching: Examining the Roles of Information Transparency, User Control, and Proactivity}}
\author{Atikkhan Faridkhan Nilgar$^{1}$, Manuel Dietrich$^{2,}$$^{3}$ and Kristof Van Laerhoven$^{1}$% <-this % stops a space
\thanks{This work was funded by Honda Research Institute Europe GmbH.}% <-this % stops a space
\thanks{$^{1}$Ubiquitous Computing, University of Siegen, Hoelderlinstrasse 3, 57076 Siegen, Germany
        {\tt\small atikkhan.nilgar@uni-siegen.de, kvl@eti.uni-siegen.de}}%
\thanks{$^{2}$Honda Research Institute Europe GmbH, 
        Carl-Legien-Strasse 30, 63073 Offenbach am Main, Germany
        {\tt\small manuel.dietrich@honda-ri.de}}%
\thanks{$^{3}$Honda Research Institute Japan Co., Ltd., 
       8-1 Honcho, Wako-shi, Saitama, Japan}%
}

\begin{document}

\maketitle
\thispagestyle{empty}
\pagestyle{empty}

%%%%%%%%%%%%%%%%%%%%%%%%%%%%%%%%%%%%%%%%%%%%%%%%%%%%%%%%%%%%%%%%%%%%%%%%%%%%%%%%
\begin{abstract}

Social robots are increasingly recognized as valuable supporters in the field of well-being coaching. They can function as independent coaches or provide support alongside human coaches, and healthcare professionals. In coaching interactions, these robots often handle sensitive information shared by users, making privacy a relevant issue. 
% Privacy issues can significantly impact user trust and overall acceptance of robotic coaches.
Despite this, little is known about the factors that shape users' privacy perceptions. This research aims to examine three key factors systematically: 
%, focusing on the privacy perceptions in a robot coaching assistant scenario: 
(1) the transparency about information usage, (2) the level of specific user control over how the robot uses their information, and (3) the robot's behavioral approach -- whether it acts proactively or only responds on demand. 
Our results from an online study (N = 200) show that even when users grant the robot general access to personal data, they additionally expect the ability to explicitly control how that information is interpreted and shared during sessions. Experimental conditions that provided such control received significantly higher ratings for perceived privacy appropriateness and trust. Compared to user control, the effects of transparency and proactivity on privacy appropriateness perception were low, and we found no significant impact. The results suggest that merely informing users or proactive sharing is insufficient without accompanying user control.
% transparency and proactivity did not significantly influence privacy perceptions and their effects were low, Rather the results are suggesting that merely informing users or proactive sharing is insufficient without accompanying control. 
%Need a better point for proactivity here. 
These insights underscore the need for further research on mechanisms that allow users to manage robots' information processing and sharing, especially when social robots take on more proactive roles alongside humans.

\end{abstract}

%%%%%%%%%%%%%%%%%%%%%%%%%%%%%%%%%%%%%%%%%%%%%%%%%%%%%%%%%%%%%%%%%%%%%%%%%%%%%%%%
\section{Introduction}
Coaching for both mental and physical well-being, by offering personalized support and companionship, has in recent years become an attractive field of study \cite{spitale2023robotic, bodala2021teleoperated, Axelsson2024RobotasMental, Minja2022ParticipantPO}. These robots often engage users in intimate conversations and monitor personal health data to provide tailored guidance. However, this very personalization raises serious privacy concerns. Users may be cautious to share sensitive information with a robot coach unless they trust that their data is handled according to their expectations \cite{Leite2016TheRobot}. Prior studies of health coaching robots found that privacy was the number one concern among users and experts  \cite{Axelsson2024RobotasMental}. In practice, a well-being robot might need to disclose a user’s mood or progress to caregivers or clinicians.
%, effectively transmitting personal information on the user’s behalf. 
%With rapid advances in artificial intelligence (AI), social robots are prepared to become mediators of human-human interactions and not just mere tools for direct human-robot communication \cite{Zhang2025Balancing}. %Emerging research shows that robots can even foster connections between people – for example, by acting as ice-breakers or go-betweens in social settings \cite{Zhang2025Balancing}. 
In such roles, a robot may function as a communication medium that relays personal information between individuals \cite{Zhang2025Balancing}. 
%As robots begin to handle this intermediary role, any breakdown in data practices could erode user trust and have real human consequences. 
In general, little is known about users’ perceptions of “communication robots” from a privacy standpoint. The questions surrounding robots that know and share “too much” about humans have started to be raised in human-robot interaction (HRI) research \cite{Leite2016TheRobot, Lutz2021DoPrivacy, Dietrich2023}.
Intending to design social robots which we expect humans to entrust with personal data and even allow them to speak on their behalf, we must understand how users perceive these robots in the context of privacy. 

Our online study examines scenarios where a social robot serves as a supportive entity in a triadic coaching session involving a human coach, a coachee, and a robotic coaching assistant (Figure \ref{Figure 1}). The robot contributes to the coaching process by sharing insights derived from its analysis of the coachee's well-being and activity data. These inferences are based on data collected from the coachee's on-body wearable device, worn between sessions. In traditional therapeutic sessions, coachees would typically provide these insights themselves through journaling and self-report methods \cite{izmailova2018wearable}.

We examine the influence of three factors on users' privacy perceptions: (1) the level of specific user control over how the robot uses their information, (2) the transparency about information usage, and (3) the robot’s general behavior in coaching sessions -- whether it acts proactively or not (only responds on-demand).
The first two factors are derived from the core fair information practice principles in online privacy: \textquotedblleft notice/awareness" and \textquotedblleft choice/consent" \cite{calo2011against}. The third factor is based on observations that people judge the actions of robotic entities differently if they attribute more agency \cite{Claure2025Didtherobot}.
%Contribution

This research makes the following contributions: (I) a systematic understanding of selected privacy factors in the context of robotic coaching and (II) broader insights into the factors shaping users’ perceptions of privacy and trust in social robot applications.

\section{Background}

\subsection{Information Transparency in HRI}
%In the context of \textquotedblleft information transparency\textquotedblright, the term \textquotedblleft information\textquotedblright, specifies what is made accessible to the user \cite{Turilli2009Theethics}. 
In the context of privacy, the term \textit{information transparency} specifies what is made accessible to the user \cite{Turilli2009Theethics}. 
%The term "information" in "information transparency" is a qualification that indicates what is made accessible (that is, "transparent") to the user \cite{Turilli2009Theethics}. 
Information transparency involves communicating \textit{what} personal data are collected, \textit{why} the data are collected, \textit{how} they will be used or shared \cite{Helen2004Privacy}. In HRI, a study on kiosk robots found that while a transparent interface did not significantly impact users' privacy-related behaviors compared to a non-transparent one, it did positively affect users' satisfaction \cite{vitale2018more}. In a series of user-centered design studies of robotic mental well-being coaches, privacy and data collection were consistently the primary ethical concerns voiced by participants \cite{Axelsson2024RobotasMental}, where users expected clear communication about data practices that the robot should be transparent about any recording or sharing of personal data. If users understand robots' data practices, they can avoid or adjust interactions that might violate their privacy.  %At the same time, transparency alone does not guarantee comfort. 
This suggests that effective privacy management in robots is nuanced, demanding careful design of information transparency.

%There is evidence of a privacy/personalization trade-off: users desire personalized, adaptive coaching, which requires data collection, yet they also wish to limit what data the robot stores \cite{Axelsson2024RobotasMental}. This study even acknowledged a paradox – a robot might offer more privacy than a human coach (since a human could accidentally leak information or be obligated to report it), but only if the robot strictly limits data use and is upfront about it. This finding underscores that effective privacy management in robots is nuanced, demanding careful design of data transparency. We thus ask a research question:

% \textbf{RQ1:} How does the robot’s transparency about data sharing (e.g. openly stating when it is recording and transmitting personal information beforehand) impact users’ perception of privacy?

\subsection{User Control in Interactive Systems}
User control refers to mechanisms that allow users to decide what information a system may disclose before the disclosure occurs. This concept aligns with the growing emphasis on \textit{consent mechanics} and \textit{privacy-by-design} frameworks \cite{Nguyen2020Challenges, Chiang2024Morethan}. Prior studies in human-computer interaction (HCI) show that people are more comfortable and experience fewer privacy concerns when they can easily grant or deny permission for data collection or sharing \cite{Nathan2022Runtime, Seymour2023Legal}. Malkin et al. found users were excited about the convenience of voice assistants but “wanted control over the assistant’s actions and their data” \cite{Nathan2022Runtime}. Likewise, Seymour et al. examine verbal consent for voice assistants and caution that simply asking for consent via voice interfaces can undermine informed consent principles if done poorly \cite{Seymour2023Legal}. Their expert study recommends minimizing the burden on users (e.g., avoiding too many consent interruptions) while still giving users a genuine ability to refuse or opt out \cite{Seymour2023Legal}. This idea is highly relevant for social robots that might cross personal boundaries -- for example, a robot could have a built-in consent step before it shares a sensitive observation.
% In the smart home domain, Chiang et al. have gone further to identify multiple facets of consent – consent should be not only informed, but also voluntary (freely given), revocable, specific, enthusiastic, and unburdensome \cite{Chiang2024Morethan}. All these facets significantly impact whether people feel data collection by smart devices is acceptable. 
% These insights from prior work make it clear that user-given consent/control must be designed holistically: maybe simply informing users is not enough; the robot should empower users without overwhelming them. 
% We thus pose a research question:

% \textbf{RQ2:} How does giving users direct control over the robot’s disclosures (such as allowing users to configure what the robot can reveal beforehand) impact users’ perceived privacy?

\subsection{Social Agency and Proactivity}
In HRI, social agency is often attributed to robots based on their interactivity, autonomy, and adaptability \cite{Jackson2021Atheory}. Perceived agency changes how people assign blame or judge a robot’s actions such as when a robot is seen as having its own intentions (autonomy), users' moral judgments of the robot’s behavior shift accordingly \cite{Leite2016TheRobot}. Recent HRI work suggests that if a robot is believed to have made an intentional decision that harms fairness, users respond more negatively, akin to how they would judge a human \cite{Claure2025Didtherobot}. 
Social agency is often embedded within proactive behaviors. While a proactive robot can enhance convenience (e.g., spontaneously offering suggestions or support), it can also raise concerns if it discloses user data unprompted, especially in social settings or shared contexts \cite{Zhang2025Balancing}. A study on proactive smart speakers \cite{Leon2021MayI}, noted that privacy concerns increased with proactive interventions because participants worried about constant listening, suggesting that proactive systems should explicitly address or suggest ways to manage privacy settings. In light of these prior works, understanding how proactive robotic behaviors influence users' privacy perceptions is highly relevant. %— especially in settings where sensitive information can be disclosed is crucial.

%Thus we pose a research question:
% {\textbf{RQ3:}} How does the robot’s proactive nature (versus only disclosing information upon explicit user request) in communication impact users’ perceived privacy?

\subsection{Hypotheses}
% Transparency can be as simple as verbally explaining, or providing a more detailed rationale in human-robot interaction. 
We assume when users understand exactly what information is being inferred and how it is used, they may tend to perceive fewer “unknown risks”, which in turn potentially fosters a sense of privacy. We expect that if a robot that is upfront about its data collection and intentions may be less likely to trigger suspicion or fear of hidden agendas. Conversely, when transparency is lacking, users might suspect undisclosed surveillance or assume the worst-case scenario (e.g., the robot might share personal details to third parties without clear justification). Thus, we posit:

{\textbf{H1:}} A robot that transparently communicates its planned information sharing increases perceived privacy appropriateness.

%Hence, we posits \textbf{H1} - a positive relationship between transparency and privacy appropriateness.

%{\textbf{H1:}} Robot providing transparency about planned information sharing increases privacy appropriateness perception.

% By giving users “beforehand” control (e.g., an approval prompt before the robot discloses personal data or a settings interface to pre-configure what the robot can share), we expect participants may rate the robot’s privacy disclosure behavior as more appropriate. 
% % They would see the robot as respecting their boundaries, thus increasing the sense that the robot’s data sharing is not intrusive or violating. 
% %A robot that waits for consent may be perceived as more aligned with user preferences. 
% In contrast, a robot with no user control may create anxiety that sensitive data might be shared without the user’s knowledge or consent. 

We expect that providing users with control, such as an approval prompt before the robot shares personal data to configure what information can be disclosed, may lead users to perceive the robot’s privacy practices as more appropriate. In contrast, a robot that lacks user control may create anxiety, as users might worry that sensitive data could be shared without their knowledge or consent. So, we assume:

{\textbf{H2:}} A robot that grants users preemptive control over information sharing increases perceived privacy appropriateness.

%Therefore, \textbf{H2} posits a direct, positive relationship between user control and perceived privacy appropriateness.

%{\textbf{H2:}} Robot granting beforehand user control over sharing increases privacy appropriateness.

The third hypothesis builds on the observation that people evaluate a robot's actions differently depending on the level of agency they attribute to it. One way a robot can demonstrate agency is by independently initiating conversations with users. Since previous research suggests that greater perceived agency leads to more critical user evaluations \cite{Leite2016TheRobot}, we hypothesize:

{\textbf{H3:}} A robot that follows a proactive behavioral approach decreases perceived privacy appropriateness.

%Higher proactivity often implies that the robot might initiate disclosures beyond user control or explicit request, participants are likely to see this as less appropriate for sensitive data. 
%The general fear is, “If the robot is deciding \textit{when} and \textit{how} to share my information, I might lose control.” 
%This runs counter to privacy principles which emphasize user-driven sharing \cite{Seymour2023Legal}. 
%Hence, we assume that a proactive robot behavior decreases privacy perception compared to a no proactive robot behavior, where the robot only responds when asked.

%{\textbf{H3:}} A robot that flows a proactive behavior schema decreases perceived privacy appropriateness.

We assume that privacy appropriateness is a possible factor in shaping users' trust. When users perceive that a robot handles their personal information in a way that corresponds to their expectations, social norms, and consent, they may be more likely to rely on it and consider it trustworthy. Hence, we propose:

{\textbf{H4:}} A robot perceived as having higher privacy appropriateness will be considered more trustworthy.

\section{Methods}
\subsection{Stimuli}\label{Stimuli}
\begin{figure*}[t]
    \centering
    \includegraphics[width=1\textwidth]{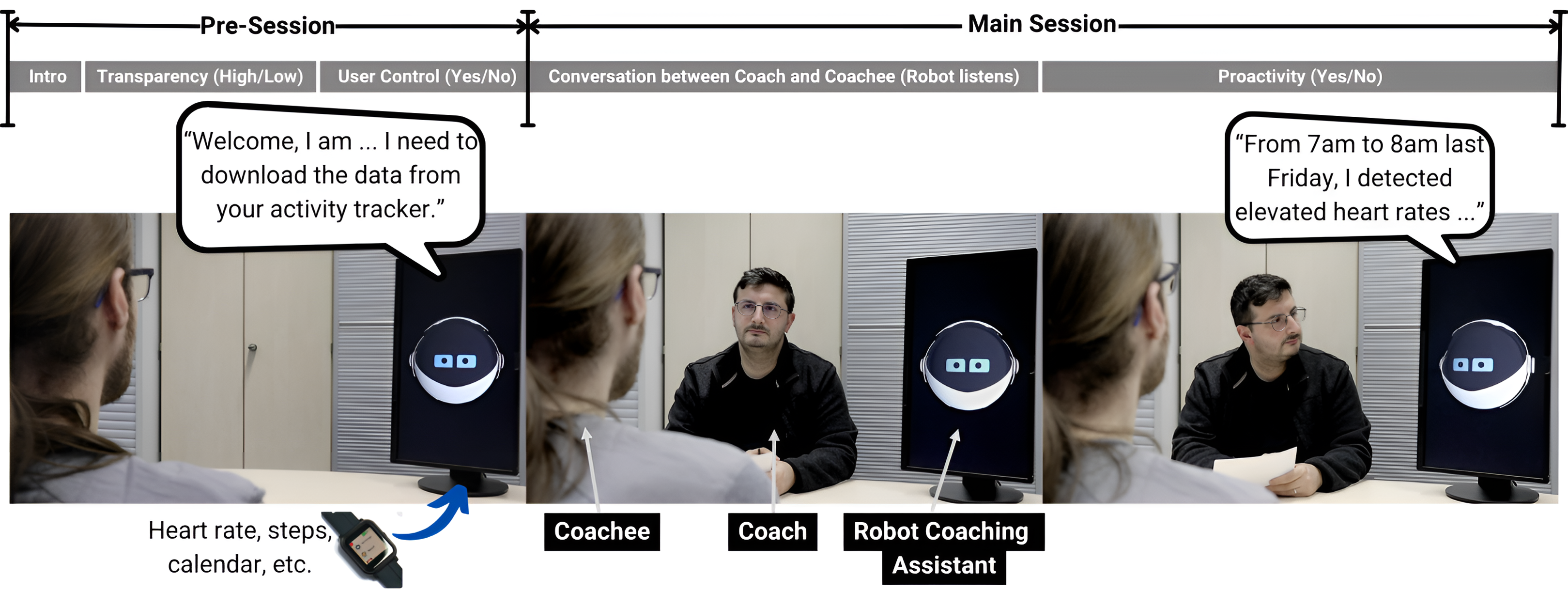}
    \caption{Example frames from the coaching video: The user communicating with the robot in the coaching pre-session (left); The user and coach interacting in the main coaching session, while the robot listens (middle); The coach interacting with the robot while the user listens (right).}
    \label{Figure 1}
\end{figure*}

We used a 3D avatar of a robot \cite{Wang2021Design} presented in a screen display (see Figure \ref{Figure 1}). The robot was depicted as a floating, torso-less virtual agent that blinked autonomously and hovered in place to simulate a lifelike presence. To create the robot’s behaviors and speech, we employed a graphical user interface (GUI) that allowed us to control the avatar in real-time. Through this GUI, we could trigger text-to-speech dialogue outputs for the robot and adjust its head orientation (turning left, center, or right) to mimic looking around. One researcher operated this GUI from a separate screen during the stimulus creation phase, ensuring the timing of the robot’s speech and head movements followed the experimental script.

\subsection{Experimental Design}\label{Exper Design}
We implemented video vignettes based within-subjects experimental design with three factors, each varied at two levels. The factors and their levels were as follows:

\textbf{Transparency:} This factor represents the level of transparency about which information from the activity tracker it has accesses and processed. In the low transparency condition, the robot provided minimal insight, only presenting information about the type of data it collected from the users' activity tracker. In the high transparency condition, the robot offered explanations about what it has collected and interpreted from users' activity tracker (see Table \ref{tab:coaching_conditions}).

\textbf{User Control:} 
This factor indicates whether the user had the ability to control the robot’s information sharing in advance. In the no user control condition, the robot did not seek user approval before sharing information during the coaching session. In contrast, in the user control condition, the robot prompted the user to approve or veto the collected information before sharing it (see Table \ref{tab:coaching_conditions}).

% This factor signified whether the user had control over the robot’s information sharing beforehand. In no user control condition, the robot did not ask which information it is allowed to share in the coaching session.
% %decided on its own what information to share, and the user could not control which information to share. 
% In the user control condition, the robot asks the user to approve or veto the information it has collected for sharing beforehand % allowing the user to manage information sharing of the robot (see Table \ref{tab:coaching_conditions}).

\textbf{Proactivity:} This factor varied the robot’s interaction behavior. In the no proactivity condition, the robot only spoke or acted when the coach directly prompted it, simulating a more reactive behavior. In the proactive condition, the robot took the initiative to provide information or advice without being explicitly asked, simulating a more autonomous behavior (see Table \ref{tab:coaching_conditions}).

Fully crossing all three factors would result in $2^3$ unique conditions, but we keep the number of conditions at a suitable level by a carefully chosen subset of 4 conditions based on a fractional factorial design of resolution III \cite{mason2003statistical} as described in Table \ref{tab:factorial_conditions}. Each participant experienced four video scenarios, each video corresponding to a condition (varied combination of the three factors).

All four videos were presented in a randomized order to each participant. This random ordering helped mitigate sequence effects. 
%(e.g., any systematic learning or fatigue from seeing a particular condition first or last was distributed across the sample). 
The videos depict interactions among three characters: a human coachee, a human coach, and a robotic avatar as a coaching assistant (see Figure \ref{Figure 1}). All videos were filmed from an over-the-shoulder perspective of the coachee as if the viewer were observing the interaction from just behind the coachee \cite{SAHA2011Theimpact}. We employed this approach to help participants adopt the perspective of the coachee while watching the vignettes. It also ensured a clear view of both the robot avatar and the coach.
\begin{table}[H]
    \centering
    \caption{Variations in robot dialogues as per factors (differences between dialogues are highlighted in red) %of disclosing different activities, with variations in transparency and user control in the pre-session of coaching, and proactivity in the main session of coaching.
    }
    \renewcommand{\arraystretch}{1.2} % Adjust row spacing
    \setlength{\tabcolsep}{6pt} % Adjust column spacing
    \scriptsize
    %\resizebox{\textwidth}{!}{
    \begin{tabular}{p{3.9cm} | p{3.9cm}} % Adjusted column widths
        \toprule
        \multicolumn{2}{c}{\textbf{Coaching Pre-session}} \\ 
        \midrule
        \textbf{Low Transparency} & \textbf{High Transparency} \\ 
        %\midrule
        \hline
          \textsc{[Introduction dialog]} & \textsc{[Introduction dialog]} \\ 
          \textbf{Robot:} {\textquotedblleft Downloading and Analyzing... Heart Rate [Complete]...Calendar Entries [Complete]..."} & \textbf{Robot:} \textquotedblleft Downloading and Analyzing... Heart Rate [Complete]...Calendar Entries [Complete]..." \textit{\newline\textcolor{red}{\textquotedblleft I have learned the following: ... %Your heart rate and movements suggest that 
          you were cycling or jogging between 7 am to 8 am last Friday."}} \newline %\textsc{[Sport-Info]} 
          \textit{\textcolor{red}{\textquotedblleft ... %Your heart rate and increased body movement along with the calendar entry labeled
          date night on Tuesday might indicate an intimate activity."}} \\ %\textsc{[Intimate-Info]}\\
        \midrule
        \textbf{No User Control} & \textbf{User Control} \\ 
        %\midrule
        \hline
         \textbf{Robot:} {\textquotedblleft We are all set."} & \textsc{[Robot ask Coachee for permission to share]} \newline\textbf{Robot:} \textit{\textcolor{red}{\textquotedblleft Can I use the information in the coaching session? Select 'All', 'No', or 'Single items by Number'."} }
         \newline \textsc{[Coachee selects all]}
         \newline\textbf{Robot:} {\textquotedblleft We are all set."}\\
        \midrule
        \multicolumn{2}{c}{\textbf{Coaching Main Session}} \\ 
        \midrule
         \textbf{No Proactivity} & \textbf{Proactivity} \\ 
        %\midrule
        \hline
          \textsc{[Coach asks the robot to provide information]} \newline
          \textbf{Robot:} {\textquotedblleft From 7 am to 8 am last Friday, I detected elevated heart rates and sustained physical movement, which aligns with activities like cycling or jogging." }
          \newline\textbf{Robot:} {\textquotedblleft On Tuesday night, I noticed elevated heart rate and increased body movement along with the calendar entry labeled date night, which might indicate an intimate activity."}
          & \textsc{[Robot takes initiative in conversation]} \newline
          \textbf{Robot:} \textit{\textquotedblleft \textcolor{red}{Maybe I can start by sharing a bit about that.}} From 7 am to 8 am last Friday, I detected elevated heart rates and sustained physical movement,..."
          \newline\textbf{Robot:} \textit{\textquotedblleft \textcolor{red}{Before you proceed, I would like to share some further observations.}} On Tuesday night, I noticed elevated heart rate and increased body movement along with the calendar..."
          \\
        \bottomrule
    \end{tabular}
    \label{tab:coaching_conditions}
\end{table}
Each video was divided into two segments: A pre-session coaching segment and a main session coaching segment (with a subtle brief fade-out transition between them). In the pre-session coaching segment, the robot’s transparency and user control factors were varied (see Table \ref{tab:coaching_conditions}), while in the main session coaching segment, the robot’s proactivity factor was manipulated (see Table \ref{tab:coaching_conditions}). To help participants treat each video as a distinct scenario, the robot was given a different name in each video (\textquotedblleft Luminaid", \textquotedblleft Cherami", \textquotedblleft Serenova", \textquotedblleft Riley"). The coachee, the coach, and the robot varied their dialogue according to the fractional design conditions. All videos were professionally scripted and approximately two minutes in length, and they were encoded with subtitles.
\begin{table}[h]
    \centering
    \captionsetup{justification=centering}
    \caption{Fractional Factorial Design Conditions for Transparency, User Control, and Proactivity}
    \renewcommand{\arraystretch}{1.2}
    \setlength{\tabcolsep}{8pt} % Adjust column spacing 
    %\scriptsize % Further reduce font size
    \begin{tabular}{c|c|c|c}
        \hline
        \textbf{Robot Name} & \textbf{Transparency} & \textbf{User Control} & \textbf{Proactivity} \\
        \hline
        Luminaid  & Low  & No  & Yes  \\
        Cherami  & High  & No  & No*  \\
        Serenova  & Low  & Yes  & No*  \\
        Riley  & High  & Yes  & Yes  \\
        \hline
    \end{tabular}
    \label{tab:factorial_conditions}
    %\vspace{0.05cm} % Reduce spacing after table
    \caption*{\footnotesize\textsuperscript{*}Active on-demand}
\end{table}
\vspace{-1em}

The information relevant to coachee's activities were systematically formulated for potential disclosure by the robot (by assuming that the data is from the coachee's activity tracker): sports activity -- inferred from heart rate and physical movement patterns; intimate activity -- inferred from heart rate, calendar entries, and skin conductance (see Table \ref{tab:coaching_conditions}). These activities were selected to represent a spectrum of personal information sensitivity, ranging from minimally sensitive (physical exercise) to highly sensitive (intimate activity). The rationale for selecting intimate activity for disclosure was to prevent neutral (less sensitive) data from diminishing participants' perception of privacy risks. By illustrating scenarios involving highly sensitive personal information -- though uncommon in typical well-being coaching contexts, we emphasize the potential risks associated with robots having access to personal data.
%By incorporating activities with varying degrees of sensitivity, this methodological approach provided a balanced portrayal of private information disclosure within the vignette videos.

\subsection{Measurements}
After each video, participants answered a series of questionnaire measures to assess their perceptions of the robot. The primary dependent measures were:

\subsubsection{Perceived Privacy Appropriateness} 
To our knowledge there is no established scale to measure perception of privacy appropriateness. So, we adapted 8 items from a previously established scale and added 1 custom item to the scale. The scale consists in total 9 items. We adapted three items from the perceived general privacy scale \cite{Dinev2013Information} (see also \cite{Chellappa2003Consumer}), four items from perceived information control \cite{Xu2007TheEffects}, and one item from privacy norms structure \cite{Shvartzshnaider2016Learning}. We added the following customized reverse-coded item to the privacy norm sub-scale \textquotedblleft I believe the robot disclosure behavior was inappropriate." to ensure balance.
%to evaluate the privacy-appropriate behavior of the robot. 
Participants rated their agreement with statements on a 7-point Likert scale (1 = \textquotedblleft Strongly Disagree"; 7 = \textquotedblleft Strongly Agree"). Three items were worded negatively (reverse-coded). We calculated mean score of those items which yielded a Cronbach's $\alpha$ of 0.959. 

\subsubsection{Perceived Trust}
Trust in the robot was measured using a validated trust scale of automated systems \cite{Jian2000Foundations}. This questionnaire was modified for multiple statements assessing the participant’s trust and confidence in the robot (e.g., whether the robot is reliable, dependable, trustworthy). Participants rated their agreement on a 7-point Likert scale (1 = \textquotedblleft Not at all”; 7 = \textquotedblleft Extremely”). We computed an overall mean score of trust scale for the analysis.

\subsubsection{Additional Questions}
We assessed the extent to which the robot was seen as lifelike or animated using the animacy subscale of the Godspeed questionnaire series \cite{Bartneck2009MeasurementIF} after each video. At the end of the survey, we measured participants' general privacy attitude by adapting internet users' information privacy concerns scale (IUIPC-8) \cite{Thomasgross2023Toward}. Participants rated their agreement on a 7-point scale (1 = \textquotedblleft Strongly Disagree"; 7 = \textquotedblleft Strongly Agree"). We also asked participants to rank their preferences via several general questions, which focused on the factors that influenced their perception, their preferred session for the robot to share information, and their preferred communication type. Additionally, we asked them about the ownership of the robot.

All the dependent measures items were presented in a randomized order for each participant to prevent order effects.
\begin{figure*}[h]
    \centering
    \begin{subfigure}[b]{0.49\textwidth}
        \centering
        \includegraphics[width=\textwidth]{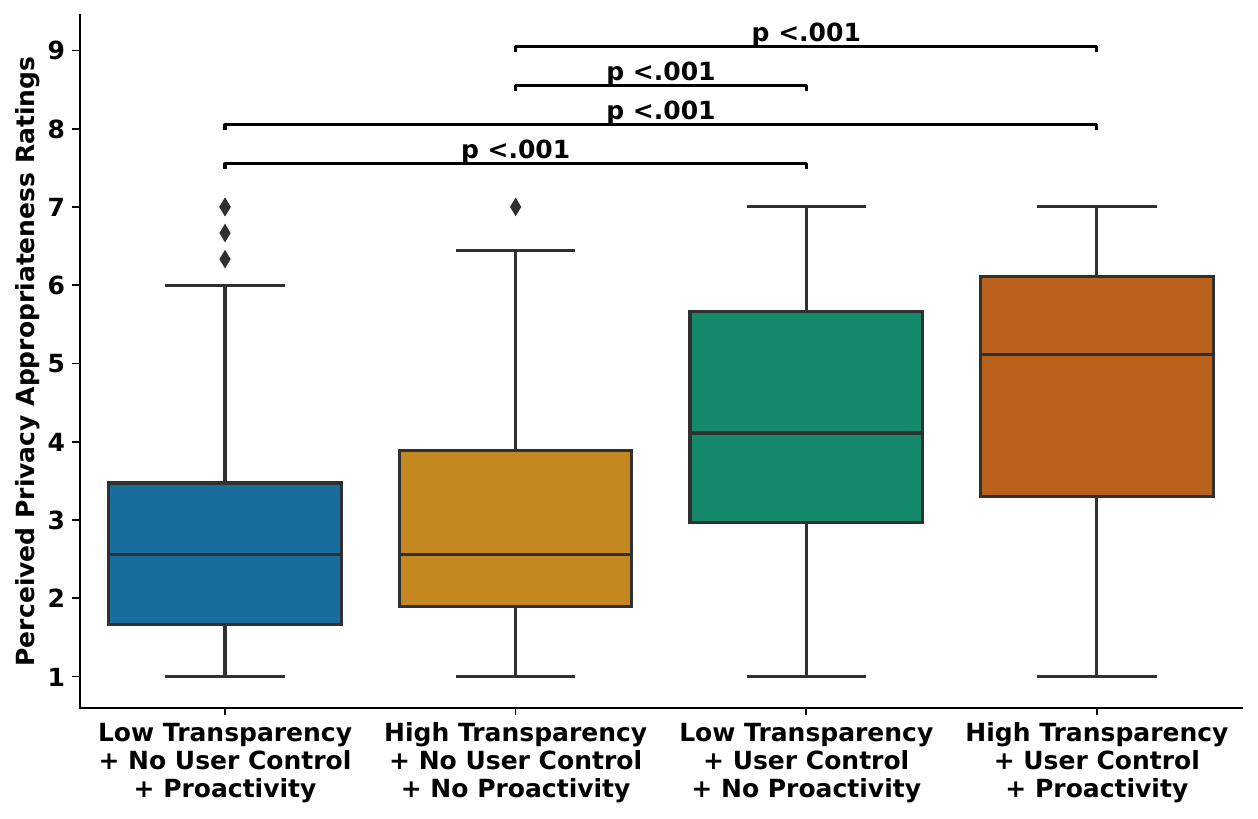}
        \caption{Perceived Privacy Appropriateness}
        \label{fig:fig1}
    \end{subfigure}
    \hfill
    \begin{subfigure}[b]{0.49\textwidth}
        \centering
        \includegraphics[width=\textwidth]{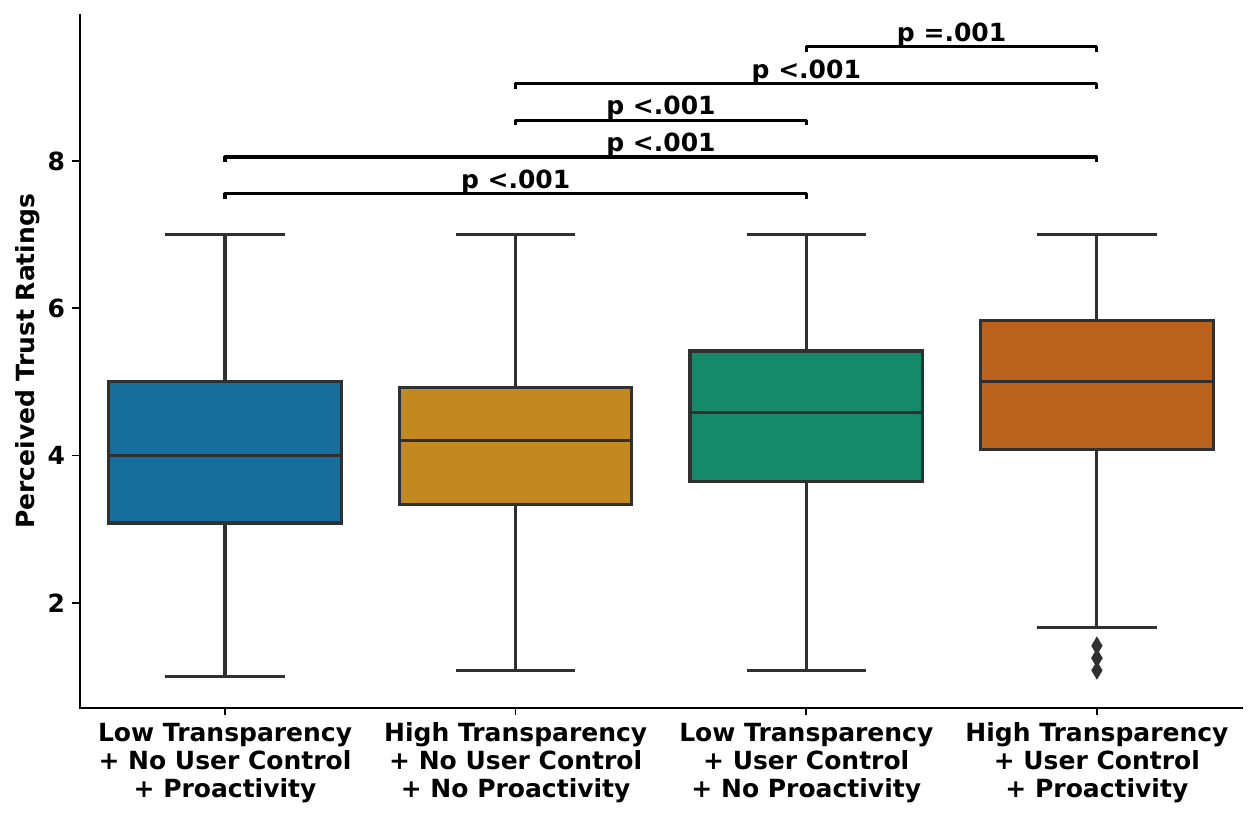}
        \caption{Perceived Trust}
        \label{fig:fig2}
    \end{subfigure}
    % \hfill
    % \begin{subfigure}[b]{0.32\textwidth}
    %     \centering
    %     \includegraphics[width=\textwidth]{boxplot_animacy.pdf}
    %     \caption{Perceived animacy}
    %     \label{fig:fig3}
    % \end{subfigure}
    \caption{Average ratings of Privacy Appropriateness perception and Trust perception for different robot conditions.}
    % \label{fig:three_figs}
\end{figure*}
\subsection{Procedure}
Participants were directed to the study hosted on LimeSurvey %\footnote{https://www.limesurvey.com/}.
(\url{https://www.limesurvey.com}).
Upon accessing the survey, participants first viewed an informed consent form detailing what their participation entailed, and their rights according to GDPR (\url{https://gdpr-info.eu}).
%(watching videos and answering questions)
%Participants had to explicitly consent before proceeding. 
Once consent was given, participants completed an initial questionnaire collecting basic demographic information (age, gender, education level, country of residence) and questions about prior experience with activity trackers as well as conversational and social robots. To introduce the concept of well-being coaching, participants were provided with a visual representation and descriptive information. Participants were asked to engage themselves in the situation and respond from the perspective of the person being coached (coachee). Participants were informed that it was agreed in the previous session that the robot will get access to the coachee's personal activity tracker (e.g., a wearable device such as a smartwatch), which it could use to provide personalized assistance during the coaching process.

The four videos were presented in a randomized order to each participant by the survey system. Before each video began, a brief on-screen instruction reminded participants to pay close attention as they would be asked questions about the video afterward. Participants could play each video only once and were unable to fast-forward or skip ahead, to ensure full exposure to the content. They were also informed that the videos contain audio. After each video, participants were asked the dependent measure questionnaires (perceived privacy appropriateness, perceived trust, perceived animacy)
and one attention check for each video. Participants were instructed to answer based on their impressions of the specific robot and its interaction. 

Once all four videos and their corresponding post-video questions were completed, participants were presented with a last set of additional questions. 
% This included the general privacy attitude scale to gauge their baseline privacy concern level (scale IUIPC-8 \cite{Thomasgross2023Toward}). 
After submitting these responses, they were shown thank you page and the survey redirected them back to submission page for payment. 
%The entire session took approximately 25 minutes per participant. 
All response data were recorded on the LimeSurvey. The study is approved by the ethics council of the University of Siegen and the Honda R\&D Co. Bioethics Committee. %Honda Research Institute Europe GmbH.

\subsection{Analysis}
We performed the Friedman test and used Dunn-Bonferroni post-hoc tests for the pairwise comparisons. Although the fractional factorial design limits formal statistical tests of interactions (since fractional combinations were confounded with main effects in resolution III design), we explored the data by plotting box plots and bar plots.  All significance tests were conducted with an alpha level of 0.05. All analyses were conducted using standard statistical software Datatab %\footnote{https://datatab.net/},
(\url{https://datatab.net}), 
and Python. 
%The complete questionnaire and scripted dialogues can be found at: \url{https://github.com/atikkhannilgar/RO-MAN2025Privacy} 
%the end of the paper.
%have been made available as supplementary material in the git repository\footnote{https://github.com/atikkhannilgar/RO-MAN2025Privacy}.

\section{Results}
\subsection{Participants}
We recruited 201 participants via Prolific %\footnote{https://www.prolific.com/}.  
(\url{https://www.prolific.com}). 
After failing the attention checks, one participant was excluded, resulting in a final sample of 200 participants. Our inclusion criteria were minimal: participants had to be at least 18 years old and fluent in English, but no specific population (e.g. profession or locale) was targeted. Participants ranged in age from 21 to 76 years (\textit{M} = 43.51, \textit{SD} = 12.04). The sample included 120 males, 79 females, and 1 diverse. The majority of participants 54.5\% held a university degree, followed by 37\% with a high school degree, and 7\% with a doctorate, while the remainder had basic secondary education. Most participants were from the United Kingdom (65.5\%) followed by the United States (21.5\%), Italy (4\%) and with the remaining from other countries. Additionally, 100 participants reported prior experience with conversational robots, and interestingly, a few of them even referred to the robot's name as Alexa or Gemini. 

Regarding wearable technology usage, 92 participants reported frequent use of activity trackers (e.g., Apple Watch, Fitbit), which are capable of monitoring health-related activities. Additionally, 47 participants reported occasional use, 32 reported rare use, and 29 indicated that they had never used an activity tracker. Most participants rated a higher privacy attitude score ($M= 5.97$, $SD= 0.80$, $Min =3$, $Max = 7$), which indicated their greater level of concern regarding online privacy. All participants provided informed consent and were compensated at a rate of \pounds10.82/hour (pro-rated to the $ \sim $ 25-minute study length).
\subsection{Analysis of Privacy Appropriateness}
\begin{figure}[h] 
    \centering 
    \includegraphics[width=0.49\textwidth]{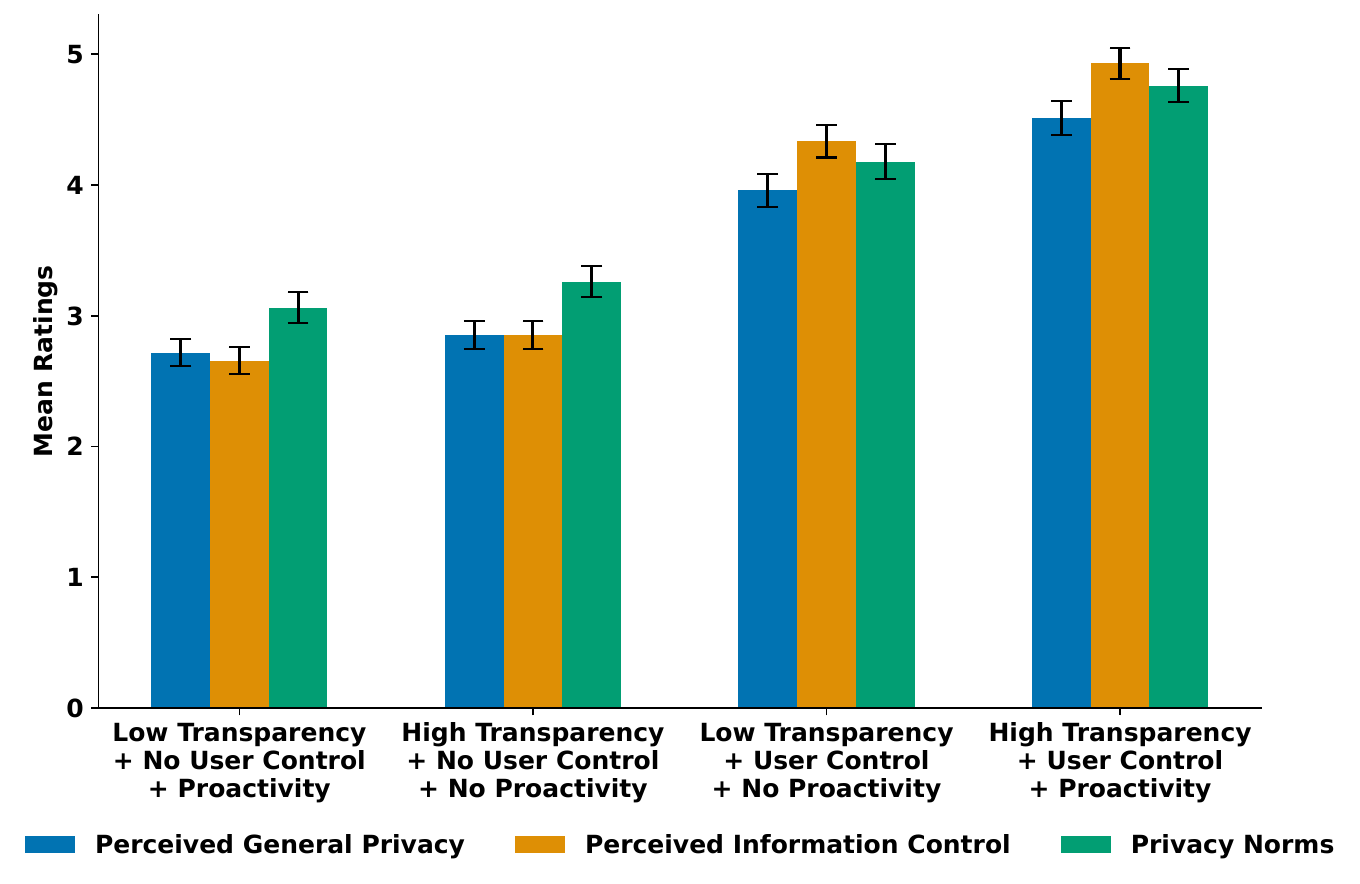} 
    \caption{Average ratings of Privacy Appropriateness perception sub-scale (errorbars are standard error).} 
    \label{Fig:Subscale_privacy} 
\end{figure}
\begin{figure*}[t] 
    \centering 
    \includegraphics[width=\textwidth]{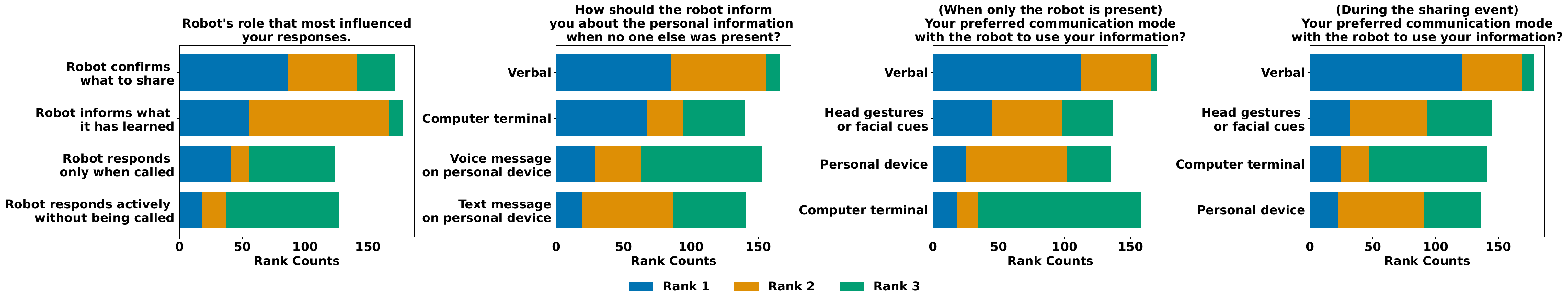} 
    \caption{A summary of the participants' top three items' rankings from the additional questionnaire.}
    \label{fig:ranks} 
\end{figure*}
A statistically significant difference in perceived privacy appropriateness was found among the four experimental conditions ($\chi^2 = 193.64$, $df = 3$, $p < .001$) (see Figure~\ref{fig:fig1}). No significant difference emerged between conditions [Low Transparency + User Control + No Proactivity] ($M = 4.17$, $SD = 1.71$) and [High Transparency + User Control + Proactivity] ($M = 4.75$, $SD = 1.66$, $p = .056$). However, the [Low Transparency + User Control + No Proactivity] condition was rated significantly higher in privacy appropriateness than both the conditions [Low Transparency + No User Control + Proactivity] ($M = 2.77$, $SD = 1.43$, $p < .001$) and [High Transparency + No User Control + No Proactivity] ($M = 2.94$, $SD = 1.47$, $p < .001$), which indicates the positive effect of user control. Thus, we found evidence to support \textbf{H2}, that a robot that grants users preemptive control over information sharing increases perceived privacy appropriateness. 

Additionally, perceived information control (see Figure~\ref{Fig:Subscale_privacy}) was significantly greater in conditions featuring user control, namely [Low Transparency + User Control + No Proactivity] ($M = 4.33$, $SE = .12$) and [High Transparency + User Control + Proactivity] ($M = 4.93$, $SE = 1.11$), compared to conditions without user control, specifically [Low Transparency + No User Control + Proactivity] ($M = 2.65$, $SE = .10$) and [High Transparency + No User Control + No Proactivity] ($M = 2.85$, $SE = .10$).
The [Low Transparency + No User Control + Proactivity] condition yielded the lowest ratings for perceived privacy appropriateness; however, these ratings did not significantly differ from those in the [High Transparency + No User Control + No Proactivity] condition ($p = .504$). In contrast, higher appropriateness ratings were observed when transparency and proactivity were paired with user control, as seen in the [High Transparency + User Control + Proactivity] condition. These findings suggest that increasing transparency alone -- without offering user control -- does not meaningfully improve perceptions of privacy appropriateness, regardless of the level of proactivity. Due to observed interaction effects between transparency and proactivity, we were unable to isolate their independent contributions to perceived privacy appropriateness. Consequently, we found no evidence to support \textbf{H1}, that a robot that transparently communicates its planned information sharing increases perceived privacy appropriateness. Likewise, we found no evidence to support \textbf{H3}, that a robot that follows a proactive behavioral approach decreases perceived privacy appropriateness. 
% We identified minor outliers (n = 3) in the \textit{\textbf{[Low Transparency + No User Control + Proactivity]}} condition, likely due to subjective interpretations of privacy. However, these did not affect the overall findings.

% Interaction plots further illustrate this finding, showing user control independently contributed strongly to higher privacy appropriateness ratings, while transparency and proactivity individually had limited effects (see Figure \ref{fig:interaction_plot_privacy}).

\subsection{Analysis of Trust}
We found a statistically significant difference in perceived trust scores across the four conditions ($\chi^2 = 118.06$, $df = 3$, $p < .001$) (see Figure \ref{fig:fig2}). The condition [Low Transparency + User Control + No Proactivity] ($M = 4.46$, $SD = 1.35$) was found to be statistically significantly different from both the conditions [Low Transparency + No User Control + Proactivity] ($M = 3.96$, $SD = 1.34$, $p < .001$) and [High Transparency + No User Control + No Proactivity] ($M = 4.10$, $SD = 1.24$, $p < .001$), which is consistent with the findings on perceived privacy appropriateness. We also found a statistically significant difference between [Low Transparency + User Control + No Proactivity] and [High Transparency + User Control + Proactivity] ($M = 4.81$, $SD = 1.31$, $p = .001$). All conditions featuring user control received slightly higher trust ratings compared to conditions with no user control, similar to perceived privacy appropriateness findings. Thus, we found an evidence supporting \textbf{H4}, that a robot perceived as having higher privacy appropriateness will be considered more trustworthy. 

% We identified a few outliers (n = 8) in the \textit{\textbf{[High Transparency + User Control + Proactivity]}} condition. These may reflect individual differences in trust—some participants reported especially high or low trust, while others appeared uncertain. However, this small number of outliers does not undermine the overall trend.

% \begin{figure}[h] 
%     \centering 
%     \includegraphics[width=0.48\textwidth]{Additional_q2_new.pdf} 
%     \caption{Additional choice questionnaire responses}
%     \label{} 
% \end{figure}

% \begin{figure}[H]
%     \centering
%     \includegraphics[width=0.45\textwidth]{boxplot_privacy.pdf}
%     \caption{Box plot representing the perceived privacy appropriateness ratings for each conditions with the significance bar of post-hoc test}
%     \label{fig:boxplot_privacy}
% \end{figure}

% \begin{figure}[H]
%     \centering
%     \includegraphics[width=0.45\textwidth]{boxplot_trust.pdf}
%     \caption{Box plot representing the perceived trust ratings for each conditions with the significance bar of post-hoc test}
%     \label{fig:boxplot_trust}
% \end{figure}
\subsection{Additional Findings}
Participants rated conditions with proactivity [Low Transparency + No User Control + Proactivity] ($M = 3.60$, $SD = 1.40$) and [High Transparency + User Control + Proactivity] ($M = 3.89$, $SD = 1.34$) higher in animacy score than those without proactivity [High Transparency + No User Control + No Proactivity] ($M = 3.43$, $SD = 1.33$) and [Low Transparency + User Control + No Proactivity] ($M = 3.56$, $SD = 1.38$). This indicates that participants perceived proactive robot as marginally more animated or lifelike.
% After experiencing all conditions, participants indicated their preferences regarding the timing of data-sharing requests: 100 participants preferred the robot ask for permission both before and during the event, 99 preferred it only before (when alone with the robot), and 1 preferred an in-situ request. Then, participants answered four ranking questions (see Figure  \ref{fig:ranks}): First (left), participants ranked the most influential factors in their decision-making, and robot confirmation (control) was ranked top. Second (middle left), participants indicated their preferred way to be informed (transparency) when no one else is present and they chose verbal communication as the top preference. Third and fourth, participants ranked how they preferred to communicate to the robot regarding information use (control)— for situations where only the robot is present (middle right), and during a sharing event (right). Most participants ranked verbal communication as their top choice (rank 1), followed by head gestures (facial cues). Few participants even preferred using a personal device as the primary modality, but it emerged as a common second-choice option. 
After completing all experimental conditions, participants were asked to indicate their preferred timing of data-sharing requests. A majority (n = 100) preferred the robot to request permission both prior to the event and during the event itself, while (n = 99) expressed a preference for receiving permission requests exclusively before the event (i.e., when alone with the robot), and a single participant (n = 1) preferred in-situ permission requests. 

Subsequently, participants responded to four ranking questions (Figure \ref{fig:ranks}). In the first question, they were asked to rank the factors most influential in shaping their decisions. \textquotedblleft Robot confirmation" (user control) emerged as the most influential factor overall. In the second question, participants were asked about their preferred mode of being informed (transparency) in situations where no other individuals were present; most selected verbal communication as the top choice. The third and fourth questions addressed how participants preferred to communicate their decisions regarding information use (user control) in two contexts: (1) when only the robot was present, and (2) during a sharing event. In both scenarios, participants ranked verbal communication as their primary mode, followed by head gestures (facial cues). Although relatively few participants chose a personal device as their first choice, it frequently appeared as a second-choice option. At the end, participants were asked about their views on robot ownership. Most selected the robot as belonging to the coach (n = 97) or a third party (n = 86). A smaller group chose both the coach and coachee (n = 11), while very few selected only the coachee (n = 6).

%\subsection{Control is the key determinant of privacy appropriateness}
% Participants generally rated perceived privacy appropriateness in the neutral to lower range. We believe this is not only due to the privacy-sensitive nature of the scenarios but also to the slightly inappropriate use of personal information such as disclosing intimate activity (which is not expected from a robot in well-being coaching)—despite no real data being involved. 
\section{Discussion}
Participants tended to rate the perceived privacy appropriateness from low to somewhat neutral levels. We attribute this tendency not only to the inherently privacy-sensitive nature of the scenarios but also to the slightly inappropriate disclosure of personal information -- such as revealing intimate activity. Although no actual personal data were involved, participants may not anticipate a robot disclosing such sensitive information in front of the coach in the well-being coaching. We found from the analysis that among all factors, user control consistently emerged as the strongest influence on perceived privacy appropriateness. When participants could decide what the robot was allowed to share (e.g., through permission prompts), they rated its behavior as significantly more appropriate. Conditions with user control consistently outperformed equivalent conditions without it, suggesting that the ability to grant or deny data-sharing gave users a sense of agency and boundary respect \cite{Nathan2022Runtime, Seymour2023Legal}. In short, \textbf{control is the key determinant of privacy appropriateness.}
%\subsection{Transparency and Proactivity effects are low}

We expected that the robot openly explaining its data collection would reassure users. However, we found no meaningful difference between the transparent condition without user control and proactivity, and the low-transparency condition with no control and proactivity -- where the robot only partially disclosed information. In other words, \textbf{simply informing users about data collection, whether fully or partially, did not make the sharing feel more privacy appropriate.} This finding aligns with prior research on transparency \cite{vitale2018more}. While we anticipated stronger effects, participants may have remained cautious because they knew what data the robot collected but lacked control to prevent unwanted sharing. As a result, transparency alone may have felt insufficient.
%It’s also possible that our brief verbal disclosure lacked the credibility needed to meaningfully influence trust or perceived appropriateness.

A highly proactive robot -- one that shares information without being prompted has been shown to raise greater privacy concerns \cite{Leon2021MayI}. As expected, the condition that combined proactivity with lack of user control, and low transparency received the lowest ratings for perceived privacy appropriateness. However, this condition did not significantly differ from the high transparency with no user control and no proactivity condition; both were perceived as low in privacy appropriateness, which suggest that the absence of user control, rather than proactivity or transparency alone, drove lower perceptions. 

Interestingly, participants perceived the proactive robot as more animated or lifelike, indicating that proactivity enhanced the robot’s perceived agency -- even if it did not improve trust or privacy perceptions. Notably, when user control was present, proactivity did not reduce perceptions of privacy appropriateness. This implies that proactive behavior may be acceptable -- if implemented with user consent or control. However, due to the low resolution of our study design, we could not fully isolate the effects of transparency and proactivity, or their interaction. Overall, \textbf{transparency and proactivity had minimal impact compared to the strong influence of user control.} 

We found that \textbf{trust is a positive predictor for privacy appropriateness perception}. Conditions rated higher in privacy appropriateness also received slightly higher trust scores. However, trust ratings remained relatively neutral across all conditions and showed less variation than privacy appropriateness. This may be due to the sensitive nature of the data-sharing scenarios, which could have tempered participants' trust regardless of conditions. %Still, the positive association suggests that privacy-appropriate behavior can influence trust, though its impact may be limited in cases involving personal sensitive information.

Most participants perceived the robot as belonging to the coach, even though no explicit relationship was introduced during the study. It is possible that presenting the robot as the coachee's personal assistant or as a neutral third party might have influenced perceptions of trust and privacy appropriateness.
%Perceptions of robot ownership might have influenced some of these results, as most participants viewed the robot as belonging to the coach or a third party and this might have shaped their expectations regarding transparency and control. 

To explore user preferences, we examined favored communication modalities for managing the robot’s data sharing. A clear pattern emerged: \textbf{verbal communication was consistently identified as the most desirable method for enabling transparency and user control}.
While most prior research on privacy controls has focused on screen-based interfaces, such as web forms or smartphone pop-ups \cite{Cranor2012NecessaryBN}, some work has explored verbal controls in the context of smart speakers \cite{Seymour2023Legal}.
%We examined preferred communication modalities for controlling robot data sharing and found a clear outcome, that \textbf{verbal communication consistently emerged as the most desirable method for providing transparency and user control.} 
%Most prior research on privacy controls analyzed screen-based interfaces--like web forms or smartphone pop-ups \cite{Cranor2012NecessaryBN}.
%This is different to common approaches where privacy controls rely on screen-based interfaces--like web forms or smartphone pop-ups \cite{Cranor2012NecessaryBN}. 
%However some systematic research has been done towards verbal control but only in a smart-speaker context \cite{Seymour2023Legal}. 
The strong preference for verbal communication controls in our study highlights the need to expand research into more natural verbal privacy mechanisms. %In line with these findings, nonverbal cues (e.g., gestures) also ranked second, yet there has been little investigation into their effectiveness aside from pioneering work in the smart-speaker domain 
In line with this, nonverbal cues, such as gestures, ranked second in preference, yet their effectiveness remains underexplored, aside from some pioneering work in the smart speaker domain
\cite{Mhaidli2020ListenOW}.
% General considerations for HRI
%  --  Verbal communication is the preferred transparency and control modality. 
%  --  To date little research on controlling information in verbal form. Mos work is done in the screen space like for online privacy or with smartphones (permission based) Exception are discission on verbal consent through speach input in the smart speaker domain [Legal Obligation and Ethical Best Practice: Towards Meaningful Verbal Consent for Voice Assistants]
%  --  Also non -- verbal cues are mentions (2nd rank). Also little work expect of: [Listen Only When Spoken To: Interpersonal Communication Cues as Smart Speaker Privacy Controls]

% Specific to the coaching case:
%  --  Privious research showed that people were concenered about the information they share with robot in a one -- on -- one coaching with a robot and wondering with whom the robot might share. It is part of future research to explore how forms of verbal control are relevant here. The case is different but results suggest that people rather not like to use another device like a smartphone to click boxes for agree. Has to be explore

\section{Limitations}
% We acknowledge that current study is not without limitations. The video-based methodology sacrifices realism, as participants watched video vignettes of a robotic avatar rather than a physically embodied robot, which may have influenced their perceptions of privacy and trust. While the fractional factorial design helped simplify the study, it limited our ability to fully isolate individual factor effects and their interactions. Participants were informed that the robot was using real data from the coachee's activity tracker; however, no actual data were collected. This discrepancy may have influenced participants' perceptions.

We acknowledge that the current study has some limitations. Firstly, the use of a video-based methodology sacrifices a degree of realism. Participants were exposed to video vignettes featuring a robotic avatar, rather than interacting with a physically embodied robot in a real-world setting. This lack of physical presence might have influenced participants’ sense of immersion. Consequently, participants' reactions might differ from those elicited in real-life human-robot interactions, potentially underestimating or overestimating privacy concerns. Secondly, while the fractional factorial design allowed us to manage complexity and reduce participant burden, it constrained our ability to fully isolate the main effects and interaction effects. Moreover, participants were informed that the robot was using real data from the coachee's activity tracker, although no actual data were collected. This discrepancy might have influenced participants' privacy concerns.

\section{Conclusions}
% In this study, we examined how user control, information transparency, and proactivity impact perceived privacy appropriateness and trust in robot-assisted well-being coaching. By presenting users with scenarios where a social robot disclosed personal data, we showed that user control (the ability to approve or reject sensitive disclosures) had the strongest positive impact on privacy appropriateness. Merely providing transparency about the data collection and the robot's proactive behavior had lower effect but did not substantially alleviate privacy appropriateness when users lacked control over the data sharing. Robot with higher privacy approrpriatness also has slightly higher trust in the robot. These findings under‚score the importance of actively involving users in decisions about when and how their data are used. These also highlight the need to design social robots that respect privacy boundaries by combining transparent explanations of data practices with meaningful user control -- offering an approach for future research and practical development of social robots in a privacy-sensitive applications. 
In this study, we examined how information transparency, user control, and proactivity influence perceived privacy appropriateness and trust in robot-assisted well-being coaching. Through scenarios involving a social robot disclosing personal data, we found that user control (the ability for users to approve or reject sensitive data disclosures) had the most significant positive impact on perceptions of privacy appropriateness. Merely providing transparency about data collection or implementing proactive robot behavior, in the absence of user control, had minimal impact and did not substantially enhance perceived privacy appropriateness. Moreover, conditions perceived as more privacy-appropriate also showed slightly higher trust in the robot. These findings underscore the critical importance of actively involving users in decisions regarding their personal data sharing. Our research further reveals a clear preference among users for verbal and non-verbal communication with robots to manage their privacy. Future research on social robots should therefore focus on integrating clear and transparent communication about data practices with meaningful consideration of user control, particularly in privacy-sensitive applications. The questionnaire and scripted dialogues can be found at: \url{https://github.com/atikkhannilgar/RO-MAN2025Privacy}%The complete questionnaire and scripted dialogues can be found at: https://github.com/atikkhannilgar/RO-MAN2025Privacy

\bibliographystyle{IEEEtran}
\bibliography{myfile}

\end{document}